\documentclass[12pt]{article}
\usepackage{amssymb,amsmath,eucal}
\begin{document}

\begin{titlepage}

                            \begin{center}
                            \vspace*{5mm}
\Large\bf{Tsallis entropy composition and the Heisenberg group}\\

                            \vspace{2.5cm}

              \normalsize\sf    NIKOS \  \  KALOGEROPOULOS $^\S$\\

                            \vspace{2mm}
                            
 \normalsize\sf Weill Cornell Medical College in Qatar\\
 Education City,  P.O.  Box 24144\\
 Doha, Qatar\\

                            \end{center}

                            \vspace{2.5cm}

                     \centerline{\normalsize\bf Abstract}
                     
                           \vspace{3mm}
                     
\normalsize\rm\setlength{\baselineskip}{18pt} 

\noindent We present an embedding of the Tsallis entropy into the 3-dimensional Heisenberg group, in order to 
understand the meaning of generalized independence as encoded in the Tsallis entropy composition property. We infer
that the Tsallis entropy composition induces fractal properties on the underlying Euclidean space. 
Using a theorem of Milnor/Wolf/Tits/Gromov,  we justify why the underlying configuration/phase space of systems 
described by the Tsallis entropy has polynomial growth for both discrete and Riemannian cases. We provide a 
geometric framework that elucidates Abe's formula for the Tsallis entropy, in terms the Pansu derivative
of a map between sub-Riemannian spaces. \\

                             \vfill

\noindent\sf  PACS: \  \  \  \  \  02.10.Hh, \  05.45.Df, \  64.60.al  \\
\noindent\sf Keywords:  Tsallis entropy, Nonextensive entropy, Heisenberg group, Abe's formula, Volume growth.  \\
                             
                             \vfill

\noindent\rule{8cm}{0.2mm}\\
   \noindent \small\rm $^\S$  E-mail: \ \  \small\rm nik2011@qatar-med.cornell.edu\\

\end{titlepage}


                                                                                 \newpage

 \normalsize\rm\setlength{\baselineskip}{18pt}

                     \centerline{\large\sc 1. \  \  Introduction}

                                                                                \vspace{5mm}

The Harvda-Charvat [1], Dar\'{o}czy [2], Tsallis [3], [4] entropy is a relatively recently introduced entropic form in Statistical Mechanics, 
which has attracted considerable attention during the last quarter of a century ([4] and references therein). 
The considerable interest in the (henceforth simply called) Tsallis entropy can be partly attributed to the recent investigations on the 
foundations of Statistical Mechanics, in particular to the investigations on the equivalence (or lack of) of the classical equilibrium ensembles [5], [6]. 
One reason motivating such investigations lies with systems with long-range interactions [7],  where in particular the concepts
of probabilistic independence, and subsequently of the additivity and extensivity of entropy have to be reconsidered [8], [4]. \\

The potential significance of the Tsallis entropy for High Energy Physics and Gravitation could turn out to  be almost as high as for Statistical Mechanics. 
After all, quantum field theories [9] are quantum theories, so they do have a statistical interpretation. Moreover and they can be seen as arising from
self-averaging of appropriate random walks, at least if one is concerned about the behavior of theories around their Gaussian fixed point [9], [10].       
It is evident that the covariant/path-integral quantization of field theories, even at zero temperature, essentially relies on the BGS entropy and its 
concomitant concepts [9], [10].  The derivation/formulation of the partition function, initially at the level of discretized variables, assumes weak 
correlations between adjacent spatial and temporal configurations [9]. This may be sufficiently adequate for weakly coupled theories with short-range 
interactions, but its use presents a huge challenge in strongly coupled theories or theories with long-range interactions as has been noticed before,
 at least since Gibbs [11].\\
 
In this vain one may try to appropriately treat gravity in the Tsallis entropy induced context, since gravity is a long-range 
interaction, from a Newtonian viewpoint. The perturbative formulation of General Relativity around the Minkowski vacuum has resulted in a 
non-renormalizable, not even Borel-summable, theory [12], [9]. This certainly calls for a different treatment in the covariant quantization of gravity: a 
potential attempt toward this goal may involve using a Tsallis entropy induced functional integral rather than the one that is currently employed.
More radical proposals certainly exist (loop gravity [13], causal sets [14], causal dynamical triangulations [15], strings/branes [16], [17] etc) 
and each has its own successes, which could possibly be enhanced by looking at them under the prism of the Tsallis, rather than the BGS entropy. 
In a more restricted context, the Tsallis entropy should be considered as a  potential candidate in the searches for the statistical origins of the black 
hole entropy [18], [19].  This is a point of tremendous interest in gravitational physics during the last four decades, which despite our best efforts has so 
far eluded a universally acceptable explanation [19].\\     

In our prior work [20], [21] we addressed the (arguably) biggest difference between the BGS and the Tsallis entropies: the way that independence and 
additivity is defined through each one of them. We concretely compared these concepts via a comparison of the composition properties of the BGS and 
the Tsallis entropies. To that end, we put the usual BGS induced additivity and the generalized  Tsallis-induced additivity side-by-side and 
established a Riemannian metric  reflecting the differences between these composition properties [21]. 
Using this approach we have been able to explain, for systems described 
by the Tsallis entropy,  why the largest positive Lyapunov exponent of the underlying dynamical system vanishes [22], derive
geometric interpretations of the nonextensive parameter [20], [21], justify the use of the escort [23], rather than the naively expected [24], probability 
distributions in applying of the maximum entropy principle [4], [24], and argue why the configuration/phase space of such systems grows at
a power-law/polynomial rate as a function of the system's number of degrees of freedom [24].\\

In the present work, we take a different path in comparing the concept of independence as encoded in the BGS and in the 
Tsallis entropy  composition properties. In Section 2,  and to keep things as simple as possible and still have composition and inversion at our disposal, 
we use simplest familiar, to us, algebraic structure possessing these operations: linear groups. To that end, we embed the Tsallis entropy into the 
the set \ $\mathfrak{U}^{3\times 3}_\mathbb{R}$ \ of \ $3\times 3$ \ real, upper triangular matrices (with units in its principal diagonal). By examining the 
corresponding Lie algebra \ $\mathfrak{u}$, \ we see that  \ $\mathfrak{U}^{3\times 3}_\mathbb{R}$ \ is a matrix realization of the 3-dimensional 
Heisenberg group.  We explicitly check that \ $\mathfrak{U}^{3\times 3}_\mathbb{R}$ \ is indeed  2-step nilpotent and provide the proof that 
\ $\mathfrak{u}$ \  is only possible other Lie algebra except the Abelian one which is nilpotent in 3 dimensions. Section 3 contains several loosely 
inter-related subsections which draw conclusions from the structure of Section 2 with various degrees of relevance to the Tsallis entropy. 
More concretely,  subsection ${\bf A}$ provides an alternative realization of the Heisenberg group via a semi-direct product construction.  
Subsection ${\bf B}$ discusses and compares the Heisenberg with the Abelian actions from the 
viewpoint of integrability. Subsection ${\bf C}$ explores the effects of the contact distribution via non-trivial holonomies and isoperimetric comparisons. 
It also addresses the fractality of the Heisneberg group. Subsection \ ${\bf D}$ \ gives a definition of the sub-Riemannian/Carnot-Carath\'{e}odory 
distance function. Subsection \ ${\bf E}$ \ discusses the bi-Lipschitz equivalence and uniqueness of the distance functions. Subsection \ ${\bf F}$ \  
discusses the dilations of the Heisenberg group. Subsection \ ${\bf G}$ \ elaborates somewhat upon the role of dilations in the Heisenberg geometry. Subsection \ ${\bf H}$ \
discusses the Pansu differential in the restricted context of interest and provides a general framework for Abe's formula of the Tsallis entropy via the 
Jackson derivative.  Subsection \ ${\bf I}$ \ establishes the power-law growth of the configuration/phase space for systems described by the Tsallis 
entropy, both for the cases of discrete and Riemannian phase spaces by using the fundamental theorem of Milnor/Tits/Wolf/Gromov on the polynomial 
growth rate of discrete groups. Section 4 concludes and presents some speculations for the implications of these constructions for Statistical Mechanics 
and Quantum Physics. \\

The present work is a considerably expanded version of [25], providing motivations, details and explanations. \\       
                                                                                     
                                                                                             \vspace{8mm}
  
 
                                                     \centerline{\large\sc 2. \ \ Tsallis entropy and the Heisenberg group}
                                                     
                                                                                           \vspace{5mm}
                                                                                                           
The Tsallis entropy $S_q$  [3], [4] is a single-parameter family of entropies, labelled by a real number \ $q\in\mathbb{R}$ \ called  
nonextensive/entropic parameter. For a set of probabilities labelled by the discrete index set \ $I$, \  it is given by
\begin{equation} 
   S_q [ \{ p_i \} ] = k_B \ \frac{1}{q-1} \left( 1- \sum_{i\in I} p_i^q \right)   
\end{equation}
where  \ $k_B$ \ is  Boltzmann's constant. In case of a probability distribution on a continuous space with corresponding probability density 
\ $\rho: \Omega \rightarrow \mathbb{R}_+$, \ 
the analogue of (1) is   
\begin{equation}
  S_q [\rho] = k_B \ \frac{1}{q-1} \left( 1 - \int_{\Omega} [\rho(x)]^q \ d\mu \right)
\end{equation}
where \ $d\mu$ \ is a Borel-regular measure on the underlying space \ $\Omega$. \ Technically, although it is sufficient for the definition of the 
Tsallis entropy  for \ $\Omega$ \ to be a space 
endowed with a measure \ $\mu$ \ in most cases of physical interest \ $\Omega$ \ is also endowed with  a metric structure. 
Probably the most commonly encountered  class of such spaces are  Riemannian manifolds \ $(M, {\bf g})$ \ for which \ $d\mu = dvol_M$ \ 
is the  Riemannian volume element uniquely associated to \ ${\bf g}$. \ It is immediate that   
\begin{equation}
    \lim_{q\rightarrow 1} = S_{BGS}
\end{equation}
where \ $S_{BGS}$ \ stands for the BGS entropy, given in the continuous case by
\begin{equation}
  S_{BGS} [\rho] = - k_B \int_M \rho(x) \log \rho(x) \ dvol_M 
\end{equation}
Henceforth we will be setting \ $k_B = 1$ \  for simplicity.\\

Statistical independence is conventionally defined  for two subsystems \ $M_1, M_2 \subset M$ \ by 
\begin{equation}  
      \rho_{M_1 \ast M_2} = \rho_{M_1} \cdot \rho_{M_2}
\end{equation}
where \ $M_1 \ast M_2$ \ indicates the combined system resulting from the interactions of \ $M_1$ \ and  
\ $M_2$. \ The Tsallis entropy (1) is not additive if we accept  this definition for independent systems, but it instead obeys
\begin{equation}
      S_q [\rho_{M_1 \ast M_2}] = S_q [\rho_{M_1}] + S_q [\rho_{M_2}] + (1-q) \  S_q[\rho_{M_1}] \ S_q[\rho_{M_2}] 
\end{equation}
To make the Tsallis entropy explicitly additive, one [26], [27] re-defines the concept of ``independence" by introducing what is essentially a 
modified Abelian group structure whose addition, reflecting (4), is given by 
\begin{equation}
    x \oplus_q y = x + y + (1-q)xy  
\end{equation}
where \ $x, y \in\mathbb{R}$. \ Our aim is to explore consequences of the generalized  definition of independence as encoded in (7). 
To that end, we start by a slight change of variables
\begin{equation}
        \tilde{S}_q [\rho ] \equiv (1-q) S_q [\rho ]
\end{equation} 
Then (6) becomes
\begin{equation} 
     \tilde{S}_q [\rho_{M_1 \ast M_2}] = \tilde{S}_q [\rho_{M_1}] + \tilde{S}_q [\rho_{M_2}] +   \tilde{S}_q[\rho_{M_1}] \ \tilde{S}_q[\rho_{M_2}] 
\end{equation} 
which amounts to inducing a modified version of the generalized addition by
\begin{equation} 
       x \ \tilde{\oplus}_q \ y = x + y + xy
\end{equation} 
We introduce an injective  map \ $\mathcal{S} : \mathbb{R} \rightarrow \mathfrak{U}^{3\times 3}_\mathbb{R}$ \ whose target space is the 
single-parameter  subset \ $\mathfrak{S}$ \ of the set of \ $3\times 3$ \ upper-triangular matrices \ $\mathfrak{U}^{3\times 3}_\mathbb{R}$ \ 
having real, equal off-diagonal elements. Concretely, 
\begin{eqnarray} 
        \mathcal{S} (x) \  = & \left( 
                                                          \begin{array}{ccc} 
                                                            1 & x & x \\
                                                            0 & 1 & x \\
                                                            0 & 0 & 1 
                                                           \end{array} \right)        
\end{eqnarray} 
and  \ $\mathfrak{S} = \{ \mathcal{S}(x), \ x\in\mathbb{R} \} $. \  Consider the usual matrix multiplication 
\begin{eqnarray} 
        \mathcal{S} (x) \ \mathcal{S}(y) \ = & \left(
                                                                  \begin{array}{ccc}
                                                                     1 & x + y & x + y + xy \\
                                                                     0 &    1    &  x + y         \\
                                                                     0 &    0    &     1
                                                                  \end{array} \right)    
\end{eqnarray} 
 which using (10) can be re-written as 
\begin{eqnarray}
       \mathcal{S}(x) \ \mathcal{S}(y) \ = & \left(
                                                                  \begin{array}{ccc}
                                                                     1 & x + y & x \tilde{\oplus}_q  y \\
                                                                     0 &    1    &  x + y         \\
                                                                     0 &    0    &     1
                                                                  \end{array} \right)      
\end{eqnarray}  
We see the map \ $\mathcal{S}$ \ is an embedding of \ $\mathbb{R}$ \ into \ $\mathfrak{U}^{3\times 3}_\mathbb{R}$ \ which 
 allows a comparison between the ordinary and the generalized addition (10) by  looking at the non-trivial off-diagonal elements of (11). 
It may be worth noticing that \ $\mathfrak{S}$ \ is not a subgroup of  \ $\mathfrak{U}^{3\times 3}_\mathbb{R}$ \ since it is not closed 
under multiplication. We will not address the question of the properties that \ $\mathfrak{S}$ \ has as a subset of \
$\mathfrak{U}^{3\times 3}_{\mathbb{R}}$ \ in this work. Instead we continue observing that as the group multiplication is probably the 
simplest form of composition, it is a highly desirable feature for our purposes. In order to have available such a composition property, 
we will be working from now on inside the ambient space \ $\mathfrak{U}^{3 \times 3}_{\mathbb{R}}$ \ which is  
\begin{eqnarray}
              \mathfrak{U}^{3 \times 3}_\mathbb{R} = & \left\{ \left(
                                                 \begin{array}{ccc}
                                                   1 & x & y \\
                                                   0 & 1 & z \\
                                                   0 & 0 & 1
                                                 \end{array} \right), \hspace{3mm} x, y, z \ \in\mathbb{R} \ \right\}    
 \end{eqnarray}  
Inside \ $\mathfrak{U}^{3 \times 3}_{\mathbb{R}}$, \ the inverse matrix of \ $\mathcal{S}(x)$ \ is given by
\begin{eqnarray}
          (\mathcal{S}(x))^{-1} \ = & \left(
                                                              \begin{array}{ccc}  
                                                                   1  & -x  & x^2 - x \\
                                                                   0  &  1  & -x  \\
                                                                   0  &  0  &  1
                                                              \end{array} \right)
\end{eqnarray}
and the group-theoretical commutator 
\begin{equation}
[\mathcal{S}(x), \mathcal{S}(y)] \equiv \mathcal{S}(x) \ \mathcal{S}(y) \  (\mathcal{S}(x))^{-1} (\mathcal{S}(y))^{-1}
\end{equation}  
is
\begin{eqnarray}   
       [\mathcal{S}(x), \mathcal{S}(y)] \ = & \left(
                                                              \begin{array}{ccc}
                                                               1 & 0 & -2xy \\
                                                               0 & 1 & 0 \\
                                                               0 & 0 & 1 
                                                              \end{array} \right)                                                             
\end{eqnarray} 
Since \ $xy \neq 0$, \ $\mathfrak{S}$ \ is not Abelian. Actually, it would be quite surprising if it were, as this would imply that the generalized 
addition (10) would essentially be the same as the ordinary addition. In turn, that would imply that the Tsallis entropy composition property 
(6)  is essentially the same as the ordinary addition. Subsequently, the sets of  axioms [28] - [30] would imply that the Tsallis entropy 
is just the BGS entropy, something which is clearly false. In this fomalism, the origin of the difference between the 
ordinary and the generalized addition (10) is exactly that  \ $xy\neq 0$ \  in (17). Now, for \ $z\in\mathbb{R}$ \ consider the commutator    
 \begin{equation}
    [\mathcal{S}(z), [\mathcal{S}(x), \mathcal{S}(y)]] =  {\bf 1}_{3 \times 3}
 \end{equation}
 where the right-hand side indicates the identity element of \ $\mathfrak{U}^{3 \times 3}_\mathbb{R}$. \. This expresses that  the subgroup
 of \ $\mathfrak{U}^{3 \times 3}_\mathbb{R}$ \ inside which \ $\mathfrak{S}$ \ is embedded,   is 2-step nilpotent.\\ 
 
 To determine the  Lie algebra \ $\mathfrak{u}$ \ of \ $\mathfrak{U}^{3 \times 3}_\mathbb{R}$ , \  we find the differential at the identity element
 \ ${\bf 1}_{3 \times 3}$, \ which gives  
 \begin{eqnarray}
    \mathfrak{u} =  & \left\{   \left(
                                         \begin{array}{ccc}
                                            0 & a & b \\
                                            0 & 0 & c \\
                                            0 & 0 & 0    
                                        \end{array} \right), \hspace{3mm} a,b,c \ \in \mathbb{R} \ \right\}
 \end{eqnarray}
We immediately verify that \ $\mathfrak{u}$ \ has as a basis 
 \begin{equation}
     X =   \left( 
            \begin{array}{ccc}
              0 & 1 & 0 \\
              0 & 0 & 0 \\
              0 & 0 & 0 
            \end{array} \right),   \hspace{7mm}
    Y =   \left(
           \begin{array}{ccc}
              0 & 0 & 0 \\
              0 & 0 & 1 \\
              0 & 0 & 0 
           \end{array} \right),   \hspace{7mm}
     Z =  \left(
            \begin{array}{ccc}
              0 & 0 & 1 \\
              0 & 0 & 0 \\
              0 & 0 & 0 
           \end{array} \right)         
 \end{equation}
The Lie algebra \ $\mathfrak{u}$ \ has Lie bracket \ $[A, B] = AB - BA, \ \ \forall \ A, B \in \mathfrak{u}$. \  The only 
non-trivial commutator is  
\begin{equation} 
     [X, Y] = Z 
\end{equation} 
with all other commutation relations being zero, as is readily seen. This is immediately recognized as the Heisenberg commutation relation
familiar from quantum Physics, where \ $Z$ \ is an element in the center of \ $\mathfrak{u}$. \ Then the group 
\ $\mathfrak{U}^{3 \times 3}_\mathbb{R}$ \  in (14) is the Heisenberg group with Lie algebra \ $\mathfrak{u}$, \ and the elements of the former  can be 
obtained by the elements of the latter  via the exponential map. It is worth noticing that the ambient space of 
\ $\mathfrak{U}^{3\times 3}_\mathbb{R}$ \  is \ $\mathbb{R}^3$ \ and that the corresponding Killing-Cartan metric is zero, as it is for any nilpotent
group. \\
 
 First, we make a comparison with the induced structure of the BGS entropy in this language. For the BGS case, we only need the ordinary addition to
 express its associated concept of ``independence". Hence, the map analogous to (11), would be
  \begin{eqnarray} 
        \mathcal{S}_E (x) \  = & \left( 
                                                          \begin{array}{ccc} 
                                                            x & 0 & 0 \\
                                                            0 & x & 0 \\
                                                            0 & 0 & x 
                                                           \end{array} \right)         
       \end{eqnarray} 
 and the general matrix group of interest, analogous to (14) is 
 \begin{eqnarray}
              \mathfrak{D}^{3 \times 3} = & \left\{ \left(
                                                 \begin{array}{ccc}
                                                   x & 0 & 0 \\
                                                   0 & y & 0 \\
                                                   0 & 0 & z
                                                 \end{array} \right), \hspace{3mm} x, y, z \ \in\mathbb{R} \ \right\}    
 \end{eqnarray}  
 which is the group of translations of \ $\mathbb{R}^3$. \ Its Lie algebra \ $\mathfrak{d}$ \ has generators 
\begin{equation}
     X_E =   \left( 
            \begin{array}{ccc}
              1 & 0 & 0 \\
              0 & 0 & 0 \\
              0 & 0 & 0 
            \end{array} \right),   \hspace{7mm}
    Y_E =   \left(
           \begin{array}{ccc}
              0 & 0 & 0 \\
              0 & 1 & 0 \\
              0 & 0 & 0 
           \end{array} \right),   \hspace{7mm}
     Z_E =  \left(
            \begin{array}{ccc}
              0 & 0 & 0 \\
              0 & 0 & 0 \\
              0 & 0 & 1 
           \end{array} \right)         
 \end{equation}
and  is, obviously, commutative.\\
 
  Second, we try to understand the difference, in the present formalism, between the ordinary and the generalized 
 additions reflecting the different views about independence and additivity that are encoded in the BGS and the Tsallis entropies. In our 
 previous work such a comparison was encoded through metrics and was made concrete by the contrast between a Euclidean (for the BGS case) 
 and a hyperbolic (for the Tsallis case) metric. The universality of the Tsallis entropy, expressed in the axioms of [28] - [30] 
 was expressed in the metric formalism via the Hadamard-Cartan theorem. In the present work, we expressed above the BGS and the Tsallis entropy 
 composition properties in a simple algebraic way by using the embedding (11). Then the difference between the BGS and the Tsallis
 entropies is reflected via the comparison of the Abelian with the Heisenberg Lie algebra (21). The universality of the Tsallis entropy, namely the 
 counterpart of the conclusions of the Hadamard-Cartan theorem in this approach, is expressed by the fact that \ $\mathfrak{u}$ \  and 
 \ $\mathfrak{d}$ \   are the only 3-dimensional 2-step nilpotent algebras over \ $\mathbb{R}$, \ as is well-known [31].  
 So, it can be claimed on rough Lie-algebraic  
 grounds, that the relation of the Tsallis to the BGS entropy is similar to that of the quantum to classical mechanics. The role of the 
 Planck constant \ $\hbar$ \ is played in the entropic case by the nonextensive parameter \ $q$. \ Elements of this analogy can also be drawn 
from  (10) which resembles the addition of probabilities in quantum Physics, with the interference term $xy$ in (10) giving rise to the novel
properties of the Quantum as compared to those Classical Physics.  \\
 
As a third point, we allude to the well-known result that \ $\mathfrak{u}$ \ and \ $\mathfrak{d}$ \  
are the only nilpotent algebras in 3 dimensions. In terms of the respective Lie groups, the statement is that the Heisenberg group \
$\mathfrak{U}^{3 \times 3}_\mathbb{R}$ \ and the group of Euclidean translations \ $\mathfrak{D}^{3\times 3}$ \ of \ $\mathbb{R}^3$ 
\ are the only two simply-connected nilpotent groups in 3 dimensions [31].  
The significance of this result  is that  it states that the Tsallis  entropy is not only the simplest, but also the only alternative to the BGS 
entropy, at this level of algebraic complexity of the underlying structures. As such, the statement of the present paragraph is universal, 
and its role in the present formalism is analogous to that of the Cartan-Hadamard theorem of the metric approach [21].  \\

                                                                                                    \vspace{8mm}


                                                         \centerline{\large\sc 3. \ \ Geometry, volume growth, derivatives}

                                                                                                    \vspace{5mm}

\noindent {\large\bf A.} \ We have repeatedly used in the previous section that  \ $\mathfrak{U}^{3\times 3}_{\mathbb{R}}$ 
\ has as underlying topological space \ $\mathbb{R}^3$ \ and, as a result, we freely switch between the Lie theoretical and the Euclidean (geometric)
viewpoints. We justify this as follows: Consider \ $\mathbb{R}^3$ \ parametrized by \ $(x,y,z)$ \ and endow it with the following inner product
\begin{equation}       
  (x_1, y_1, z_1)\cdot (x_2, y_2, z_2) \ = \ \left( x_1 + x_2, \  y_1 + y_2, \  z_1 + z_2 + \frac{1}{2} (x_1 y_2 - x_2 y_1) \right) 
\end{equation}
This inner product  defines the (sub-Riemannian) Carnot-Carath\'{e}odory metric tensor \ ${\bf g}_{CC}$. \ We observe that left-translations/left-
multiplications preserve the horizontal distribution \ $\mathcal{H}$ \ which is the subset of $\mathfrak{U}^{3\times 3}_{\mathbb{R}}$ generated by \
$X, Y$ \ in (20).   This can be explicitly checked for each one of the \ $X, Y, Z$ \ as follows: Consider a left-translation 
\ $L_{(a,b,c)}$, \ with \ $(a,b,c) \in\mathbb{R}^3$, \ acting on \ $(x,y,z)$, \ namely
\begin{equation}  
     L_{(a,b,c)} (x,y,z) = (a,b,c)\cdot (x,y,z) = \left( a+x, \ b+y, \ c+z + \frac{1}{2} (ay-bx) \right)
\end{equation}
The corresponding differential is given by the Jacobian matrix
\begin{equation}
    (dL)_{(a,b,c)} = \frac{ \partial L_{(a,b,c)} (x,y,z) }{\partial (x,y,z)} = \left( 
                                                                                                                            \begin{array}{ccc}
                                                                                                                              1 & 0 & 0   \\
                                                                                                                              0 & 1 & 0  \\
                                                                                                                              -\frac{b}{2} & \frac{a}{2} & 1
                                                                                                                            \end{array} \right)
\end{equation}
Consider \ $X$ \ as a column matrix having coordinate components \ $X = (1, 0 , -\frac{y}{2})$ \  and  
\begin{equation}
   (dL)_{(a,b,c)} X = \left( 
                                      \begin{array}{ccc}
                                      1 & 0 & 0   \\
                                      0 & 1 & 0  \\
                                     -\frac{b}{2} & \frac{a}{2} & 1
                                      \end{array} \right)                                                                                                                                                           
                                  \left(
                                      \begin{array}{c}
                                         1 \\
                                         0 \\
                                         -\frac{y}{2}
                                      \end{array} \right)        
\end{equation}
giving 
\begin{equation}
(dL)_{(a,b,c)} X = \frac{\partial}{\partial x} + \left( -\frac{b}{2} - \frac{y}{2}\right) \frac{\partial}{\partial z}
\end{equation}
Moreover 
\begin{equation}
       X\circ L_{(a,b,c)} = \frac{\partial}{\partial x} - \frac{b+y}{2} \frac{\partial}{\partial z}
\end{equation}
giving 
\begin{equation}
(dL)_{(a,b,c)} X = X \circ L_{(a,b,c)} 
\end{equation}
which is what we wanted to establish for \ $X$. \ Working in a similar manner, we find
\begin{equation}  
(dL)_{(a,b,c)} Y = \frac{\partial}{\partial y} + \frac{a+x}{2} \frac{\partial}{\partial z} = Y \circ L_{(a,b,c)}
\end{equation}
 and 
 \begin{equation}
 (dL)_{(a,b,c)} Z = \frac{\partial}{\partial z} = Z \circ L_{(a,b,c)}
 \end{equation} 
 Therefore, indeed each left translation (26) is an isometry of \ ${\bf g}_{CC}$.\
  Consider now the map \ $\Psi : \mathbb{R}^3 \rightarrow \mathfrak{U}^{3\times 3}_\mathbb{R}$ \ given by 
\begin{equation}
   \Psi:  (x, y, z) \mapsto \left(
                                                \begin{array}{ccc}
                                                    1 & x & z + \frac{1}{2}xy \\
                                                    0 & 1 & y \\
                                                    0 & 0 & 1 
                                                \end{array} \right)
\end{equation}
This is a Lie group isomorphism between \ $\mathbb{R}^3$ \ endowed with the inner product (26) and \ $\mathfrak{U}^{3\times 3}_\mathbb{R}$ \ 
endowed with the usual matrix multiplication, as can be readily checked. Under this isomorphism, we see that the basis vectors of $\mathbb{R}^3$ map 
precisely to the basis matrices of (20). 
It is worth mentioning that this inner product (26) is strongly reminiscent of the warped product in the solvable group construction that lead to the effective 
hyperbolic metric in [21]. That was subsequently used in [22], [24]  for uncovering, with the addition of simplifying assumptions, 
several properties of systems described by the Tsallis entropy. This is not a coincidence as all nilpotent groups, such as the Heisenberg group, are 
obviously solvable, hence the above sub-Riemannian construction expresses the same properties, as the hyperbolic metric construction [21], 
of the Tsallis entropy composition, even if seen from a different viewpoint.\\


\noindent{\large\bf B.} \ It is worth noticing that \ $X, Y$ \ alongside \ $[X, Y]$ \ span the whole \ $\mathfrak{u}$. \ This is unusual, from a Euclidean 
viewpoint, where all three vectors \ $X, Y, Z$ \ are needed to determine the Lie algebra of translations. The difference is  that for the
 case of Euclidean translations, the corresponding Lie algebra (23) is commutative, hence all the commutators between the basis vectors vanish.
For the Euclidean case, the basis vectors \ $X_E, Y_E$ \ can be chosen to lie in the ``horizontal" plane $xy$, by choosing as in (24)
\begin{equation}
    X_E = \frac{\partial}{\partial x}, \hspace{10mm} Y_E = \frac{\partial}{\partial y}, \hspace{10mm} Z_E = \frac{\partial }{\partial z}
\end{equation}
By contrast, due to the nature of the Heisenberg algebra, when $X, Y$ are horizontal, namely orthogonal to the center of \ $\mathfrak{u}$ \ 
the vector \ $[X,Y]$ \ will lie in the \ $z$, \ the ``vertical", direction. As above, the horizontal distribution of planes generated by $X,Y$ is denoted by 
\ $\mathcal{H}$. \ Choose  
\begin{equation}
   X =  \frac{\partial}{\partial x} - \frac{y}{2} \frac{\partial}{\partial z}, \hspace{10mm}
             Y =  \frac{\partial}{\partial y} + \frac{x}{2}  \frac{\partial}{\partial z}, \hspace{10mm}
                     Z = \frac{\partial}{\partial z}
\end{equation}
whose commutators satisfy the Heisenberg algebra (21). Dually, and equivalently, one may use co-frames
and express the Heisenberg condition (21) via the vanishing of the contact form
\begin{equation}
       \Xi \ = \  dz - \frac{1}{2} (xdy - ydx)   
\end{equation}  
where \ $d$ \ indicates the exterior derivative. Then the kernel of \ $\Xi$ \ 
\begin{equation}
      \ker \Xi \ = \left\{ (\beta_1, \beta_2, \beta_3) \in\mathbb{R}^3 \ : \ \beta_3 = \frac{1}{2} \left( \beta_1 d\beta_2 -  \beta_2  d\beta_1 \right) \right\}
\end{equation}
determines locally \ $\mathcal{H}$, \ which is the base space of the submersion \ $\mathbb{R}^3 \rightarrow \mathbb{R}^2$. \\
  
In the case of Euclidean translations, the vectors \ $X_E, Y_E$ \ (35) span a plane which is 
the leaf of a foliation of \ $\mathbb{R}^3$ \ by ``horizontal" planes parallel to the span of \ $X_E, Y_E$ \ 
and which are parametrized by \ $z$. \
By contrast, in the Heisenberg case, the vectors \ $X, Y$ \ (36) span, locally, a plane which does not form a foliation of \ $\mathbb{R}^3$. \ 
One way to express this contrast is via Frobenius theorem: in the Euclidean case, the plane distribution is integrable, whereas in the 
Heisenberg case, the contact distribution is (maximally) non-integrable. 
More generally, the commutators of a nilpotent Lie algebra result in enough, linearly independent, vectors that span their 
whole linear space. \\


\noindent{\large\bf C.} \ Consider a curve \ $\gamma : [0, 1] \rightarrow \mathbb{R}^3$ \
starting at the origin \ $\gamma (0) = 0 \in \mathbb{R}^3$ \ whose co-framing belongs to \ $\ker \Xi$ \  i.e. which its everywhere tangent  
to the horizontal distribution \ $\mathcal{H}$. \ 
Moreover assume that  the endpoint of this curve lies on the z-axis $\gamma(1) = (0, 0 , z)$. Consider its projection on \ $\mathcal{H}$ \ and call it \
$\tilde{\gamma}$. \ Then, the area \ $A$ \ of the disk \ $D$ \ that \ $\tilde{\gamma}$ \ encloses \ is given by  
\begin{equation} 
    A = \int_D  dA = \int_D  dx \wedge dy
\end{equation}
where \ $\wedge$ \ indicates the exterior/wedge multiplication of differential forms. We observe that \ $dA$ \ is the differential of an element 
of \ $\ker \Xi$. \ Combining (38) and Stokes' theorem we find that  
\begin{equation} 
      A =  \int dz = z
\end{equation} 
Notice that $z$ is the Euclidean distance between \ $\gamma(0)$ \ and \ $\gamma(1)$. \
Therefore the non-integrable distribution (38) gives rise to non-trivial holonomies of curves. 
This formalism has been extensively used in describing adiabatic phases acquired by states of a quantum system (Berry phase, Hannay 
angles etc) [32], in the description of mechanical systems with constraints [33] and in classical  field theory in models involving gauge fields coupled 
to scalars or fermions [9]. In the current context, this result can be interpreted as stating that the generalized addition (10) is akin to introducing 
a multiplicative non-integrable phase to the composition of the probabilities of the two interacting sub-systems \ $M_1, M_2$ \ of (6). This phase 
expresses in a different way, the generalized concept of independence, usually encoded by the long-range temporal and spatial correlations 
of the distributions  \ $\rho_{M_1}$ \ and \ $\rho_{M_2}$ \ (6). The entanglement of subsystems expressed through such a non-integrable phase 
is one of the conjectured features of systems described by the Tsallis entropy.\\

Moreover, looking at the process leading to (40) one can also infer the following [34]: due to the isoperimetric inequality in the  plane
\begin{equation}
  A \leq \frac{1}{4\pi} \ l_{CC}^2(\tilde{\gamma})
\end{equation}         
where \ $l_{CC}$ \ stands for the length of its argument with respect to the sub-Riemannian metric \ ${\bf g}_{CC}$. \ Because \ $\tilde{\gamma}$ \ 
is the projection of \ $\gamma$ \ on \ $\mathcal{H}$, \ we have
\begin{equation}
  A \leq \frac{1}{4\pi} \ l_{CC}^2(\gamma)
\end{equation}
According to (40) the Euclidean distance \ $d_E$ \ between the endpoints of \ $\gamma(0)$ \ and \ $\gamma(1)$ \ is \ $A$, \  and given that \ 
$l_{CC}(\gamma) = d_{CC}$ \ between the endpoints we see that 
\begin{equation}  
    \sqrt{4\pi d_E}   \   \leq   \  d_{CC}
\end{equation}
On the other hand 
\begin{equation}
        d_{CC} \ \leq \  h(\gamma) \sqrt{4 \pi d_E}
\end{equation}  
where \ $h(\gamma)$ \ is some function depending on the endpoints of \ $\gamma$. \ The last equation expresses the fact that the
metric is Euclidean on \ $\mathcal{H}$. \ We conclude that in the z-direction, perpendicular to \ $\mathcal{H}$ \ or the complement $\ker \Xi$, the 
sub-Riemannian metric \ ${\bf g}_{CC}$ \ behaves like the square root of the Euclidean metric \ ${\bf g}_E$, \ whereas along \ $\mathcal{H}$ \ it behaves 
like \ ${\bf g}_E$. \ Hence the metric \ ${\bf g}_{CC}$ \ is highly anisotropic from a Euclidean perspective.\\

To explore consequences of this,  consider two small cubes, \ $I_E \in (\mathfrak{D}^{3\times 3}, {\bf g}_E)$ \ and 
 \ $I_{CC} \in (\mathfrak{U}^{3\times 3}_{\mathbb{R}}, {\bf g}_{CC})$. \  Due to the previous arguments  the areas of their faces along \ $\mathcal{H}$ \  
 scale in both cases as \ $t^2$ \ where \ $t>0$ \ is a scaling parameter. In the direction orthogonal to \ $\mathcal{H}$ \ the length of the Euclidean 
 cube \ $I_E$ \ scales as \ $t$, \ as expected. By contrast, in the same direction, the cube \ $I_{CC}$ \ has length that scales as  \ $t^2$ \  as can be seen 
 from (43), (44). \ So, we arrive at the scaling relations 
\begin{equation} 
    Vol (I_E) \sim t^3
\end{equation}
and
\begin{equation}
   Vol (I_{CC}) \sim t^4
\end{equation}
Then, essentially by definition, the Hausdorff dimension of \ $(\mathfrak{D}^{3\times 3}, {\bf g}_E)$ \  is 3 and that of  
\ $(\mathfrak{U}^{3\times 3}_{\mathbb{R}}, {\bf g}_{CC})$ \ is 4.
We observe that the underlying topological space in both of these cases is \ $\mathbb{R}^3$ \ 
whose topological dimension is 3. Hence the space \ $(\mathfrak{U}^{3\times 3}_{\mathbb{R}}, {\bf g}_{CC})$ \ is a fractal [35], [36]. By that we mean 
exactly a space whose Hausdorff dimension is strictly greater than its topological dimension [35] . Note, in passing, that the opposite inequality is 
impossible [37] and that the definition of  fractals [35] does not require their Hausdorff dimensions to be non-integers (see however [36]).\\

 We arrived in a fractal structure for    
\ $(\mathfrak{U}^{3\times 3}_{\mathbb{R}}, {\bf g}_{CC})$ \ by restricting the admissible curves to be tangent to the non-integrable distribution 
\ $\mathcal{H}$. \ Then the distance between any two points, defined as the infimum of all curves joining them, is quite different for the 
sub-Riemannian metric \ ${\bf g}_{CC}$ \ than for the Euclidean metric \ ${\bf g}_E$. \ By this realization we come back to the origins of 
the Tsallis entropy: the introduction of the Tsallis entropy relied on the multi-fractal formalism [3], [4]. Naturally it should encode some of its 
properties. Probably the simplest such manifestation is that the concept of ``independence" adopted by the Tsallis entropy, 
via the generalized addition (10), gives  rise to a fractal structure on the \ $\mathbb{R}^3$ \ when compared to the Euclidean structure induced by the 
BGS entropy. The above considerations pertaining to the dimension are very special cases of the ball-box theorem and of the general 
Bass/Mitchell/Gershkovich/Nigel-Stein-Waigner [33], [34]  formula giving the homogeneous  dimension of  sub-Riemannian 
spaces at their regular points. \\ 


\noindent{\large\bf D.} \ As a result of the isomorphism \ $\Psi$\ (34), one can equivalently  define the sub-Riemannian metric \ ${\bf g}_{CC}$ \ on 
\ $\mathfrak{U}^{3\times 3}_\mathbb{R}$ \ 
to be the Riemannian metric with respect to which the basis vectors (36) are orthonormal [33], [34]. 
It turns out that \ ${\bf g}_{CC}$ \ is the restriction of  the Euclidean metric \ ${\bf g}_E$ \ on the horizontal distribution \ $\mathcal{H}$ \ 
spanned by \ $X, Y$ \ (36). In the Lie algebra language the argument is re-expressed as follows: by inverting (36) we get  
\begin{equation}
     \frac{\partial}{\partial x} = X + \frac{y}{2} Z, \hspace{10mm} \frac{\partial}{\partial y} = Y - \frac{x}{2} Z
\end{equation}
Let  
\begin{equation}
W = w_1\frac{\partial}{\partial x} + w_2 \frac{\partial}{\partial y} + w_3 \frac{\partial}{\partial z}
\end{equation}
After using (48), we have 
\begin{equation}
                  W = w_1 X + w_2 Y + \left( w_3 - \frac{1}{2}(xw_2 - y w_1) \right) Z
\end{equation}
and assuming that \ ${\bf g}_E(W, \cdot) \in \ker \Xi$ \ we get from (49) that the coefficient of  \ $Z$ \ vanishes, thus giving  the sought after
\begin{equation}
        W = w_1 X + w_2 Y
\end{equation}

The Carnot-Carath\'{e}odory distance between two elements \ $A, B$ \ of \ $\mathfrak{U}^{3\times 3}_\mathbb{R}$ \ 
is defined as the length \ $l_{CC}(A, B)$ \ of the shortest horizontal curve \ $\gamma: [0 ,1] \rightarrow \mathcal{H}$ \  joining \ $A, B$, \ 
namely by 
\begin{equation}
       d_{CC} (A, B) = \inf_{\gamma}  \{ l_\gamma (A, B) \}    
\end{equation}
One can verify that any two points on \ $\mathfrak{U}^{3\times 3}_\mathbb{R}$ \ can be connected by such a horizontal curve of 
finite length with respect to \ ${\bf g}_{CC}$ \ as follows [34]: consider the one parameter subgroups \ $R_X, \ R_Y$ \ of right translations    
of \ $\mathfrak{U}^{3\times 3}_\mathbb{R}$ \ generated by \ $X, Y$ \ in (20). These two single parameter families are tangent to \ $\mathcal{H}$. \  
Since \ $X, Y$ \ are the generators under the commutator of \ $\mathfrak{u}$, \ the subgroups \ $R_X$, \ $R_Y$ \ they generate, also generate 
the whole of \ $\mathfrak{U}^{3\times 3}_\mathbb{R}$. \ Therefore any two points of \ $\mathfrak{U}^{3\times 3}_\mathbb{R}$ \ can be connected by 
a piecewise smooth curve, whose every smooth piece is generated either by \ $R_X$ \ or by \ $R_Y$. \     
This is clearly impossible in \ $(\mathbb{R}^3, {\bf g}_E)$. \  It does however happen in 
\ $(\mathfrak{U}^{3 \times 3}_{\mathbb{R}}, {\bf g}_{CC})$ \  because  \ $\mathcal{H}$ \ allows for 
curves that ``turn and twist" sufficiently to make it possible. The general statement substantiating this pictorial explanation, applicable to  
any sub-Riemannian space with vector fields having iterated Lie brackets spanning their whole tangent bundles, 
is due to Chow and Rashevskii [33], [34].  
From the sub-Riemannian viewpoint, the Tsallis entropy composition (7), (10) essentially amounts to the existence of a horizontal 
distribution \ $\mathcal{H}$.\\ 


\noindent{\large\bf E.} \ It may also worth briefly commenting on the uniqueness of \ ${\bf g}_{CC}$. \ Looking in the above construction, 
we see that the Tsallis entropy 
composition property (7), (10) actually determines the map (11). Upon embedding the image of  (11) into (14), 
we could use any metric we wished on (14) to determine the properties of interest, as the construction itself does not instruct  
us how to choose a ``natural", from the physical viewpoint, metric.
One can even attempt to work outside the realm of Riemannian spaces, with Finslerian metrics, for instance. Then it is not obvious that the
results obtained, certainly the ones pertinent to the concept of ``independence" of subsystems of a physical system, would not be dependent of such a 
construction. In short, apart from its familiarity and simplicity, why is the use of \ ${\bf g}_{CC}$, \ or equivalently, of the inner product (25)
optimal for drawing conclusions about physical systems? To make things somewhat manageable, we constrain ourselves to sub-Finslerian metrics, 
in analogy with sub-Riemannian metrics, namely to distance functions \ $d_1$ \ and \ $d_2$ \ calculated along horizontal distributions which are 
arising from (Banach) norms \ $||\cdot ||$. \ Consider two such sub-Finslerian metrics \ $||\cdot ||_1$ \ and \ $||\cdot ||_2$ \ giving rise to \ $d_1$ \ and 
\ $d_2$ \ respectively. The physical properties of the systems that we have in 
mind do not change, as will also be seen in the sequel, if the distances \ $d_1$, \ $d_2$ \ do not change all that much, that is if such distances change, 
but in a uniformly controlled manner.\\

Using (35) and (38), this  is expressed precisely by stating that the physical 
properties of interest  associated with the concept of ``independence" do not change if the two distance functions \ $d_1, \ d_2$ \ are bi-Lipschitz equivalent, 
namely if there are constants \ $c_1>0, \ c_2>0$ \ such that           
\begin{equation}
   \frac{1}{c_2} \ d_2(x,y) \leq d_1(x,y) \leq c_2 \ d_2(x,y), \hspace{5mm} \frac{1}{c_1} \ d_1(x,y) \leq d_2(x,y) \leq c_1 \ d_1(x,y), 
                  \hspace{5mm} x,y\in \mathbb{R}^3 
\end{equation}
However, it turns out that  all sub-Finslerian spaces of interest in this subsection are bi-Lipschitz equivalent, 
so it makes no difference for the physical aspects of the system under study which particular metric we 
choose to work with. As such we choose the most flexible and familiar ones, namely the sub-Riemannian 
ones arising from inner products such as  (25).\\ 


\noindent {\large\bf F.} \ Another point  worth noticing is that \ $\mathfrak{u}$ \ has a simple dilation property mirroring the corresponding 
symmetry of \  $\mathbb{R}^3$. \ The dilations \ $\tilde{\delta}_t, \ t\in\mathbb{R}_+ $ \  of \ $\mathfrak{D}^{3\times 3}$ \ are a single parameter family of 
diffeomorphisms given by
\begin{equation}                  
                                     \tilde{\delta}_t  \left(
                                                 \begin{array}{ccc}
                                                   x & 0 & 0 \\
                                                   0 & y & 0 \\
                                                   0 & 0 & z
                                                 \end{array} \right)  =   \left(
                                                 \begin{array}{ccc}
                                                   tx & 0 & 0 \\
                                                   0 & ty & 0 \\
                                                   0 & 0 & tz
                                                 \end{array} \right)   
\end{equation}
and they are isometries of \ $(\mathfrak{D}^{3\times 3}, {\bf g}_E)$. \ 
In \ $\mathfrak{U}^{3\times 3}_\mathbb{R}$, \ due to the existence of the generalized addition in (12), 
such a simple behavior of the metric under dilations is not possible. One observes however that the more general, graded re-scalings 
\begin{equation}
   \delta_t  \left(
                                                 \begin{array}{ccc}
                                                   1 & x & z \\
                                                   0 & 1 & y \\
                                                   0 & 0 & 1
                                                 \end{array} \right)  =   \left(
                                                 \begin{array}{ccc}
                                                   t & tx & t^2 z \\
                                                   0 & t & ty \\
                                                   0 & 0 &  t
                                                 \end{array} \right)   \end{equation}
result in the following simple-looking behavior for the sub-Riemannian distance function 
\begin{equation}
       d_{CC} (\delta_t(A), \delta_t(B)) \ = \ t \ d_{CC} (A, B), \hspace{8mm} A,B \in \mathfrak{U}^{3\times 3}_\mathbb{R}
\end{equation}
This can be checked explicitly from (54). We have to pay attention to that \ $d_{CC}$ \ is calculated along curves that are tangent to \ $\mathcal{H}$ \ 
which re-scale linearly under \ $\delta_t$. \ Hence the corresponding distance function \ $d_{CC}$ \ also re-scales linearly under the above dilations.
The fact that the central element scales quadratically with \ $t^2$ ,\ and not linearly, is not really important in this argument as its behavior is 
already taken care of via the bracket-generating condition (21). On the other hand, using the dilations (54), we can re-establish the previously reached
result (46) that  the small balls of \ ${\bf g}_{CC}$ \ have volumes scaling as \ $t^4$ \ hence that the Hausdorff dimension of \ 
$\mathfrak{U}^{3\times 3}_\mathbb{R}$ \ is actually \ $4$ \ instead of the naively expected, at a first glance, \ $3$.  \\


\noindent{\large\bf G.} \  Consider the Heisenberg group \ $\mathfrak{U}^{3\times 3}_\mathbb{R}$ \ endowed with a left-invariant 
Riemannian metric \ ${\bf g}$. \ Let \
\begin{equation}
     {\bf g}_t = \frac{1}{t} \ \delta_t ^{\ast} ({\bf g})
\end{equation}
where \ $\delta_t ^{\ast}$ \ is the dilation of the metric induced from (54). The corresponding distance function \ $d_t$ \ converges to \ $d_{CC}$ \    
for \ $t\rightarrow\infty$ \ in the pointed Gromov-Hausdorff sense [38], [39].
 Moreover ($\mathfrak{U}^{3\times 3}_\mathbb{R}, \ d_t$) \ is isometric to \
($\mathfrak{U}^{3\times 3}_\mathbb{R}, \ \frac{1}{t} d$). \  Hence the asymptotic cone [40], [41] of \ $\mathfrak{U}^{3\times 3}_\mathbb{R}$ \  
endowed with the Riemannian distance function \ $d$ \ is \ $\mathfrak{U}^{3\times 3}_\mathbb{R}$ \ endowed with the sub-Riemannian distance 
function \ $d_{CC}$. \ This is a geometric way of looking at the generalized addition (7), (10)  induced by the Tsallis entropy. 
From the Euclidean/Riemannian perspective, which is expressed via the usual addition, the role of (7), (10) it two-fold: it gives a rise to a non-integrable
distribution \ $\mathcal{H}$ \ essentially due to the quadratic term of (7), (10). Naturally the space remains Riemannian for any value of the dilation 
parameter. In the large dilation limit though, the curves that are perpendicular to \ $\mathcal{H}$ \ are suppressed and distance calculations are 
only allowed along curves that are tangent to \ $\mathcal{H}$. \ This expresses geometrically that the difference between the usual and the generalized 
additions may be of little consequence for small values of numbers that are summed, but  it becomes progressively important and eventually dominant 
as these numbers increase. Eventually the inequality in the degrees of the terms of the generalized addition makes it more akin to multiplication than to 
addition. In other words, if someone thinks about addition of entropies for independent systems for the BGS case, this person should think in 
multiplicative terms  when discussing the concept of ``independence" for systems described by the Tsallis entropy.\\     

The dilations, being isometries, also reflect the self-similar behavior of \ ($\mathfrak{U}^{3\times3}_\mathbb{R}, {\bf g}_{CC}$) \ at different scales. 
Given the origin of the Tsallis entropy in the multi-fractal formalism, this behavior  should not come as a surprise. This self-similarity at different scales 
shows that the Heisenberg group behaves in a roughly similar way to the many other fractals encountered in Physics, almost all of which have 
non-integer Hausdorff dimension [35], [36]. By considering the limits, the dilations  allow us to explore the local sub-Riemannian geometry of the 
Heisenberg group by examining its asymptotic cone on which one would expect to be able to use several simplifications in the models under study. 
The fact that dilations provide isometries between the different cones has an additional implication: the Heisenberg group is purely 2-unrectifiable. 
Notice that it is not purely 1-unrectifiable as most of curves in it, not ``escaping" to infinity, have finite length. Being 2-unrectifiable [42] means that 
all rectifiable sets of Hausorff dimension 2, namely all sets of finite area, have 2-dimensional Hausdorff measure zero with respect to the sub-
Riemannian metric ${\bf g}_{CC}$. \ This expresses in a precise way that 
the overwhelming majority of 2-dimensional sets in the Heisenberg group \ $\mathfrak{U}^{3\times 3}_\mathbb{R}$ \ are unrectifiable. Moreover the 
ones that are rectifiable, are insignificant in number (a rare exception), by comparison. This is in accordance with our intuition where fractals are so 
convoluted that (depending on the particular fractal) lengths, areas etc are infinite when measures are considered at a very fine scale. 
Pictorially and for a comparison, it may worth seeing this 
observation as a multidimensional extension of the von Koch snowflake curve, any segment of which has infinite length hence any segment of which is 
purely 1-unrectifiable [36]. Again, this behavior induced by the composition property of the Tsallis entropy, equivalently by the 
generalized definition of independence, can be traced back to the origins of the Tsallis entropy in the multii-fractal formalism [3], [4]. 
Thus once more we are brought back  full circle to the set of ideas and structures behind the introduction of the Tsallis entropy. \\   
  
   
\noindent{\large\bf H.} \  Given the unrectifiable nature of subsets of the Heisenberg group discussed the previous subsection, it may be of interest to 
examine to what extent structures induced by the Tsallis entropy are differentiable. This allows us to inquire to what extent first order calculus can be
developed for such sets in order to probe their infinitesimal behavior. It also allows us to wonder about the degree of smoothness of these structures
and far more importantly, especially for Quantum Gravity, to what extent such smoothness is a fundamental or an emergent phenomenon [14], [15], 
[42] - [44].\\

One can always wonder what is a local analogue of the ordinary derivative which somehow reflects the 
composition properties of the Tsallis entropy. Such a derivative has been constructed  for Heisenberg (actually for stratified/
Carnot) groups in [45]. Pansu actually showed that a tangent map of a Lipschitz map between sub-Riemannian spaces exists, is unique and is a 
group homomorphism of the tangent cones which is equivariant with the corresponding dilations. This happens for almost all points of the Heisenberg 
group, except possibly for a set of Hausdoff measure zero. Hence there is a unique differentiable structure associated 
with a Lipschitz map which is best adapted to the dilation properties of the tangent cones to the Heisenberg group [45]. 
To be concrete, let us consider a map \  $\phi: \mathfrak{U}^{3\times 3}_\mathbb{R} \rightarrow \mathbb{R}^n$ \ from the Heisenberg group to the 
Euclidean space. Let the left translations on either group be indicated by abuse of notation by \ $\mathcal{L}$  \ and let the corresponding 
dilations be indicated by  (54) and  (53) respectively. The Euclidean space \ $\mathbb{R}^n$ \  is seen here as a 1-step nilpotent 
(Abelian) group. The corresponding tangent map \ $P \phi$ \ given by Pansu's theorem  [45] at \ $A\in\mathfrak{U}^{3\times 3}_\mathbb{R}$ \ is 
\begin{equation}     
       (P \phi) (A) \  = \  \lim_{t\rightarrow\infty} \ \tilde{\delta}_{t^{-1}} \circ \mathcal{L}^{-1}_{(\phi_A)} \circ \phi \circ \mathcal{L}_A \circ \delta_t \ A 
\end{equation} 
 Let's be a bit more concrete for a couple of special cases of interst to the BGS and Tsallis entropies. 
 First consider the map \ $\varphi: \mathfrak{D}^{3\times 3} \rightarrow \mathfrak{D}^{3\times 3}$ 
 between the diagonal subgroups of $\mathbb{R}^{3\times 3}$ which represent Euclidean translations. 
 This map is generally given by
 \begin{equation}
   \varphi:     \         \left( 
                                \begin{array}{ccc}      
                                       x & 0 & 0 \\ 
                                       0 & y & 0 \\
                                       0 & 0 & 0  
                                \end{array}    \right)
                                    \    \mapsto \  \left(
                                                      \begin{array}{ccc}
                                                           \varphi (x) & 0 & 0 \\
                                                           0 & \varphi (y) & 0 \\
                                                           0 & 0 & \varphi (z) 
                                                       \end{array} \right)
 \end{equation} 
 Consider the matrix \ $\bar{A}\in\mathfrak{D}^{3\times 3}$ \ in a neighborhood of which the Pansu derivative will be calculated and the 
 corresponding variation matrix to be \ $\bar{X}\in\mathfrak{D}^{3\times 3}$. \ These have the form   
 \begin{equation}
                     \bar{A} =  \left( 
                                \begin{array}{ccc}      
                                       a & 0 & 0 \\ 
                                       0 & b & 0 \\
                                       0 & 0 & c  
                                \end{array}    \right),
                                    \   \hspace{8mm} \  \bar{X} =  \left(
                                                                              \begin{array}{ccc}
                                                                                   x & 0 & 0 \\
                                                                                   0 & y & 0 \\
                                                                                   0 & 0 & z 
                                                                              \end{array} \right)
\end{equation} 
with all entries being in \ $\mathbb{R}$. \  The dilation (53) of \ $\bar{X}$ \ gives  
\begin{equation}
              \delta_t  \bar{X} = \left(
                                                 \begin{array}{ccc}
                                                    tx & 0 & 0 \\
                                                    0 & ty & 0 \\
                                                    0 & 0 & tz 
                                                 \end{array} \right)
\end{equation}
and the left translation of \ $\bar{A}$ \ by \ $\delta_t \bar{X}$ \ has image under \ $\varphi$ \ given by
\begin{equation}
          \varphi (\bar{A} + \delta_t\bar{X}) \ = \  \left(
                                                 \begin{array}{ccc}
                                                   \varphi (a + tx) & 0 & 0 \\
                                                        0   & \varphi (b+ty) & 0 \\
                                                        0  &   0   & \varphi (c+tz)
                                                 \end{array} \right)
\end{equation}
Since \ $\mathfrak{D}^{3\times 3}$ \ is an Abelian group under addition,
\begin{equation}
     [\varphi (\bar{A})]^{-1} = \left(
                                                  \begin{array}{ccc}
                                                      -\varphi (x) & 0 & 0 \\
                                                               0        & -\varphi (y) & 0\\
                                                               0        &  0  & -\varphi (z)
                                                  \end{array} \right)  
\end{equation}
which gives 
\begin{equation}
     \tilde{\delta}_{t^{-1}} \left( [\varphi (\bar{A})]^{-1}  \varphi (\bar{A} + \delta_t \bar{X})  \right) \ = \
                                          \left( 
                                              \begin{array}{ccc}
                                                 \frac{\varphi (a+ tx) - \varphi (a)}{t} & 0 & 0 \\
                                                           0  &  \frac{\varphi (b+ty) - \varphi (b)}{t} & 0\\
                                                           0   &   0  & \frac{\varphi (c+tz) - \varphi (c) }{t}       
                                             \end{array} \right)
\end{equation}
Upon taking the  limit \ $t\rightarrow 0$ \ we find
\begin{equation}
         (P \varphi ) (\bar{A}) \ = \ \left(
                                               \begin{array}{ccc}
                                                   \frac{\partial \varphi (a)}{\partial x} & 0 & 0 \\
                                                         0 & \frac{\partial \varphi (b)}{\partial y} & 0 \\
                                                        0 &  0  & \frac{\partial \varphi (c)}{\partial z}
                                               \end{array} \right)
\end{equation}
So, we see that in the particular case of (64) its Pansu derivative is the Euclidean gradient, a fact compatible with 
our intuition about derivatives.\\

As a second example, consider  the map \ 
$\Phi: \mathfrak{U}^{3\times }_\mathbb{R} \rightarrow \mathfrak{D}^{3\times 3}$ \ given by
\begin{equation}  
      \Phi:     \         \left( 
                                \begin{array}{ccc}      
                                       1 & x & z \\ 
                                       0 & 1 & y \\
                                       0 & 0 & 1  
                                \end{array}    \right)
                                    \    \mapsto \  \left(
                                                      \begin{array}{ccc}
                                                           \Phi (x) & 0 & 0 \\
                                                           0 & \Phi (y) & 0 \\
                                                           0 & 0 & \Phi (z) 
                                                       \end{array} \right)                
\end{equation}
with 
\begin{equation}
        \widetilde{A} = \left(
                              \begin{array}{ccc}
                                  1 & a & c   \\  
                                   0 & 1 & b  \\ 
                                   0 & 0 & 1  
                              \end{array} \right),     \hspace{8mm}
                                                                         \widetilde{X} =  \left(
                                                                                      \begin{array}{ccc}
                                                                                          1 & x & z \\
                                                                                          0 & 1 & y \\
                                                                                          0 & 0 & 1
                                                                                      \end{array}  \right)
\end{equation}
in analogy with (59). Then following the steps of the previous example, we have the dilations (54) 
which give
\begin{equation}
      \Phi ( \widetilde{A} \delta_t \widetilde{X} ) \ = \  \left(
                                                     \begin{array}{ccc}
                                                         \Phi (a + tx) &         0          &           0 \\
                                                             0               & \Phi (b+ty)  &           0\\    
                                                            0               &          0          &       \Phi (c+tz)
                                                     \end{array} \right)
\end{equation}
Moreover, since \ $\mathfrak{D}^{3\times 3}$ \ is Abelian
\begin{equation}
     [\Phi (\widetilde{A})]^{-1} \  =  \ \left(
                                                          \begin{array}{ccc}
                                                                 -\Phi (a) & 0 & 0 \\
                                                                      0 & -\Phi (b) & 0 \\
                                                                      0 & 0 & -\Phi (c)
                                                          \end{array} \right)
\end{equation}
Let's confine ourselves to a neighborhood of the identity in \ $\mathfrak{U}^{3\times 3}_\mathbb{R}$. \ This is not a 
loss of generality as we can recover the behavior of any quantity in the neighborhood of any point of \
$\mathfrak{U}^{3\times 3}_\mathbb{R}$ \  by acting by left translations on a neighborhood of the identity. 
Assume then that \ $a = b = c = 0$ \ in (67). Then (68) gives \ $\Phi (a) = \Phi (b) = \Phi (c) = 0$, \ which results in  
\begin{equation}
         \tilde{\delta}_{t^{-1}} ([\Phi (\widetilde{A})]^{-1}  \Phi (\widetilde{A} + \delta_t \widetilde{X}) ) \ = \ 
                          \left(  
                                   \begin{array}{ccc}
                                       \frac{\Phi (tx)}{t}  & 0 & 0 \\
                                           0 & \frac{\Phi (ty)}{t} & 0 \\
                                           0 &  0 & \frac{\Phi (tz)}{t}   
                                   \end{array} \right)
\end{equation}
The Jackson derivative is defined for a function \ $h:  \mathbb{R} \ \rightarrow \ \mathbb{R}$ \ by 
\begin{equation}
      \frac{D}{dx} \ h(x) \ = \ \lim_{t\rightarrow 1} \ \frac{h(tx) - h(x)}{tx - x}
\end{equation}
when the limit exists. Given this definition, we see that the Pansu derivative of the map \ $\Phi$ \ at the origin of \ 
$\mathfrak{U}^{3\times 3}_\mathbb{R}$ \ is given by the Jackson derivatives of its entries 
\begin{equation}
       (P \Phi ) (\widetilde{A}) \ = \ \left(
                                                           \begin{array}{ccc}
                                                                \frac{D}{\partial x} \Phi (0) & 0 & 0 \\
                                                                     0 & \frac{D}{\partial y} \Phi (0) & 0 \\
                                                                     0 & 0 & \frac{D}{\partial z} \Phi (0) 
                                                           \end{array} \right)
\end{equation}
It is an interesting observation  by  Abe [46] that the BGS entropy can be written as
\begin{equation} 
       S_{BGS} \ = \ - \frac{d}{dx} \left( \sum_{i\in I} p_i^x \right) \bigg|_{x=1}
\end{equation}
and that a similar relation holds for the Tsallis entropy where the ordinary has been substituted
by the Jackson derivative 
\begin{equation}
      S_q \ = \ - \frac{D}{dx} \left(  \sum_{i\in I} p_i^x \right) \bigg|_{x=1} 
\end{equation}
The sub-Riemannian framework established above for the Tsallis entropy clarifies why (73) is 
the natural counterpart of  (72). It also shows the important, but distinct, role of dilations in both cases. 
Moreover it provides a unified geometric framework for both the ordinary 
and the Jackson derivatives which are seen as special case of the Pansu derivatives of maps 
between sub-Riemannian spaces. \\ 


\noindent{\large\bf I.} \ Another conclusion of the Heisenberg group construction given above is the following: 
recall [38] - [40]  that a Riemannian manifold \ $M$ \ has polynomial (power-law) growth 
when the volume \ $Vol: \ M \rightarrow \mathbb{R}_+$ \ of all balls \ $B_r(x), \ x\in  M$ of radius \  $r$  \  does not increase faster than some 
power of $r$, namely, when there is an \ $p\in\mathbb{R}$ \ such that
\begin{equation}    
   Vol (B_r(x)) \leq  r^p 
\end{equation}
Naturally, all Riemannian manifolds \ $M$ \ are infinitesimally Euclidean, so for small \ $r$, \ we have
\begin{equation}
   \lim_{r\rightarrow 0} Vol (B_r(X)) \leq \mathrm{const} \  r^n
\end{equation}
where \ $n$ \ is the topological dimension of \ $M$. \ So, for such \ $M$, \ the above definition may become non-trivial when one considers 
the limit \ $r\rightarrow\infty$. \ This amounts to  taking the asymptotic viewpoint and wondering how the manifold looks at large scales 
(from large distances).  This asymptotic viewpoint does not distinguish between $M$ itself and a set of points in $M$ having similar 
large-scale geometric characteristics. This geometric indistinguishability is encoded in the concept of quasi-isometry [38]  - [41].
A map \ $f: M_1 \rightarrow M_2$ \   between two metric spaces $M_1$, $M_2$ with corresponding distance functions \ $d_1$ \ and 
\ $d_2$ \ is a quasi-isometry, if there are constants \ $c_1 >0, \ c_2 > 0$ \ such that 
\begin{equation}    
    \frac{1}{c_1} \ d_1(x,y) - c_2 \leq \ d_2(x, y) \  \leq c_1 \ d_1(x,y) + c_2
\end{equation}
for all \ $x ,y \in M_1$. \ In words, this means the following: we discretize \ $M_1$  \ and \ $M_2$  \ by choosing nets with respect to $d_1$ and $d_2$ 
respectively. By doing so, we ignore all small-scale structures in \ $M_1$ \ and \ $M_2$. \ This is the role the constant $c_2$ in (76). The  
two nets are bi-Lipschitz equivalent. This makes precise 
that \ $d_1(x,y)$ \  should not be distorted too much by \ $f$. \ The maximal distance distortion
is determined by the constant \ $c_1$. \ In this, large scale, treatment we are interested in quasi-isometrically invariant features of the underlying 
structures of \ $M_1$ \ and \ $M_2$. \   In the case of interest to us, we pick as a net in \ $\mathfrak{U}^{3\times 3}_\mathbb{R}$, \ the discrete 
Heisenberg group \ $\mathfrak{U}^{3 \times 3}_\mathbb{Z}$ \  having integer entries
\begin{eqnarray}  
\mathfrak{U}^{3 \times 3}_\mathbb{Z} = & \left\{ \left(
                                                 \begin{array}{ccc}
                                                   1 & x & y \\
                                                   0 & 1 & z \\
                                                   0 & 0 & 1
                                                 \end{array} \right), \hspace{3mm} x, y, z \ \in\mathbb{Z} \ \right\}    
\end{eqnarray}
This is a finitely generated group, and mirroring the case of real coefficients, it is also nilpotent. Choose a set of generators \ $\mathfrak{T}$ \  of  \ 
$\mathfrak{U}^{3\times 3}_\mathbb{Z}$ \  by picking
\begin{equation}
    T_1 =   \left( 
            \begin{array}{ccc}
              1 & 1 & 0 \\
              0 & 1 & 0 \\
              0 & 0 & 1 
            \end{array} \right),   \hspace{7mm}
    T_2  =   \left(
           \begin{array}{ccc}
              1 & 0 & 0 \\
              0 & 1 & 1 \\
              0 & 0 & 1 
           \end{array} \right)                            
 \end{equation}
 The generating set \ $\mathfrak{T}$ \ should be made symmetric by including in it  the inverses of its generators   which are 
\begin{equation}
    (T_1)^{-1} =   \left( 
            \begin{array}{ccc}
              1 & -1 & 0 \\
              0 & 1 & 0 \\
              0 & 0 & 1 
            \end{array} \right),   \hspace{7mm}
    (T_2)^{-1}  =   \left(
           \begin{array}{ccc}
              1 & 0 & 0 \\
              0 & 1 & -1 \\
              0 & 0 & 1 
           \end{array} \right)                                   
 \end{equation} 
Express any element \ $\tilde{U}$ \ of \ $\mathfrak{U}^{3\times 3}_\mathbb{Z}$ \ in terms of the generators of the symmetrized set \ $\mathfrak{T}$ \  
by the word and reduce 
\begin{equation}
   \tilde{U} = (T_1)^{k_1} (T_2)^{k_2} \cdots (T_2)^{k_i}
\end{equation}
Such a representation of an element of \ $\mathfrak{U}^{3\times 3}_\mathbb{Z}$ \  in terms of elements of \ $\mathfrak{T}$ \ is not unique. 
The length of the above word representing $U$ is by definition $|k_1| + |k_2| + \ldots + |k_i|$. The norm \ $||U||$ \ of \  $U$ \  is the minimum length of 
all words that express  \ $U$ \ in terms of elements of \ $\mathfrak{T}$ [39]. \ Let \ $B_r$ \ indicate the closed ball with respect to the generating set \ 
$\mathfrak{T}$ \ having as center the identity element \  $\mathbf{1}_{3\times 3}$ \ of \ $\mathfrak{U}^{3 \times 3}_\mathbb{Z}$ \ and radius \ $r$, \ 
namely
\begin{equation}
    B_r = \{ U\in\mathfrak{U}^{3\times 3}_\mathbb{Z}: \ ||U|| \leq r \}
\end{equation}
The volume of \ $B_r$ \ is defined with respect to the counting measure, hence it is the cardinality of the underlying set. The group \ 
$\mathfrak{U}^{3\times 3}_\mathbb{Z}$ \ has polynomial growth if there are two constants \ $c>0$, \ $d>0$ \ such that
\begin{equation}   
         Vol \ B_r \leq c r^d
\end{equation} 
One can prove that the property of polynomial growth does not depend on the choice of the generating set \ $\mathfrak{T}$. \ indeed, changing the 
generating set will only affect the value of $c$. The value of the exponent $d$ depends only on the group $\mathfrak{U}^{3\times 3}_\mathbb{Z}$ 
and not on the choice of the generating set \ $\mathfrak{T}$. \ Hence  any change of the 
generating set will amount to a quasi-isometry of \ $\mathfrak{U}^{3\times 3}_\mathbb{Z}$ \ but the polynomial growth property is invariant 
under quasi-isometries. A fundamental theorem of Milnor [47], Wolf [48], Tits [49], Gromov [39] (see also the more recent [50], [51])
states that a finitely generated group has polynomial growth if and only if it is virtually 
nilpotent, namely if  contains a nilpotent subgroup of finite index. In our case of interest,  the ``virtual" adjective is not actually needed as 
\ $\mathfrak{U}^{3\times 3}_\mathbb{Z}$ \ is nilpotent itself. Not too surprisingly,  the growth rates of the balls in \ 
$\mathfrak{U}^{3\times 3}_\mathbb{R}$ \ and in \ $\mathfrak{U}^{3\times 3}_\mathbb{Z}$ \ are equivalent. Hence the growth 
rate of the balls in \ $\mathfrak{U}^{3\times 3}_\mathbb{R}$ \ is also polynomial. Since these structures reflect the composition 
property  of the Tsallis entropy. At this point we make the very strong assumption, that such a behavior is present at the level of phase space dynamics, 
and it is not emergent from the statistics. Then this polynomial/power-law growth of balls in the Heisenberg group \ 
$\mathfrak{U}^{3\times 3}_\mathbb{R}$ \ actually reflects a similar behavior of the volume in the configuration/phase space of the underlying 
microscopic system  whose statistical properties are encoded by the Tsallis entropy. 
Seen from a different viewpoint, this polynomial behavior is ascribed to strong spatial and temporal 
correlations that constrain the growth rate of the volume function in the configuration/phase space of the system. This is a result 
that was reached in [52], [53] for binary systems and was generalized in [24] for Riemannian manifolds. The present conclusion, 
reached through a different path when compared to the previous works, has the advantage that it is equally applicable to both cases 
of discrete and continuous systems at the same time. \\


\noindent{\bf\large J.} \ An important question that someone can ask is how different actually the Tsallis entropy is from the BGS entropy. 
From the above constructions it appears that this amounts to asking how different are the resulting sub-Riemannian from the corresponding 
Riemannian spaces. The answer is 
clearly a matter of quantifying ``far" and ``close". To do so in metric way we use bi-Lipschitz maps. 
For two spaces that are bi-Lipschitz, 
the distances are not distorted too much, so such spaces can be considered as equivalent if someone observes them from some distance.  
So the question is whether there a  bi-Lipschitz map between the Heisenberg group \ $\mathfrak{U}^{3\times 3}_\mathbb{R}$ \ and \ 
$\mathbb{R}^3$. \ By using Pansu's generalization of Rademacher's theorem to mappings between sub-Riemannian spaces, Semmes [54] 
answered the question in the negative. He proved [54] that there is no bi-Lipschitz embedding from an open subset of a Heisenberg group to a 
Euclidean space. In that sense the underlying geometry of the Tsallis entropy is quite different from that of the Euclidean space which is 
induced by the composition of the BGS entropy. Intuitively the picture is clear: if such a bi-Lipschitz map existed then it would have to be almost 
everywhere differentiable. Its Pansu differential would have to be a group homomorphism. The approximating maps would have to be also 
bi-Lipschitz due to the homogeneity of the blow-up cones. Therefore such maps would have to be injective which is impossible because the whole 
center of the Heisenberg group would be mapped to a point in the Euclidean space in order to have a group homomorphism. 
This result can be extended  when the target is \ $\mathbb{R}^n$ \ for any \ $n\in\mathbb{N}$ \ and can be seen as the lack of an analogue of 
Nash's embedding theorem for sub-Riemannian spaces.  
As a by-product one sees that the Heisenberg group is purely 1-unrectifiable as was also noticed in subsection {\bf G} above. \\

                                                                                                  \vspace{8mm}
 

                                                             \centerline{\large\sc  4. \ \ Discussion and outlook}              

                                                                                                \vspace{5mm}
         
We presented above a sub-Riemannian construction encoding the effects of the generalized addition induced by the Tsallis entropy 
composition property (7), (10). We showed how features of this construction can be used to justify known properties of the Tsallis entropy.
We would like to point out that all the mathematical facts we used, except the particulars pertaining to the Tsallis 
entropy, are well-known  even classical to practitioners of Geometry. What we have attempted in this work, was by barely scratching the surface of 
the existing knowledge of sub-Riemannian/Carnot-Carath\'{e}odory  geometry, to provide the connections with and to show how 
some of the the Tsallis entropy properties can be organized from a  sub-Riemannian perspective. 
It turned out that some previously known, but not obviously inter-connected, facts about the Tsallis 
entropy fit neatly in the sub-Riemannian framework, which provides an effective geometric language that expresses in a unified way many 
aspects of the Tsallis entropy. Moreover, the Heisenberg group perspective highlights the universality and uniqueness features that distinguish 
the Tsallis entropy from other entropic functionals that have been developed over the last few decades [4].  \\  
    
One can ask whether there is any connection between the sub-Riemannian construction presented here and the the hyperbolic map \ 
$\tau_q$ \ used in [20], [21], [22], [24]. The answer to that is affirmative and is provided by the boundary construction of (Gromov)-hyperbolic 
spaces and the visual/Tits metrics on it, for non-compact symmetric spaces of rank 1 [38] - [40], [55]. 
We will elaborate upon these constructions, point out their connections with the Tsallis entropy and explore their potential physical 
implications  in a future work.\\
    
                                                                                    \vspace{5mm}
 
 
                                                      \centerline{\large\sc Acknowledgement} 

                                                                                     \vspace{3mm}

\noindent We are grateful to A.J. Creaco, initial discussions with whom motivated this line of investigation.                                                                            
 We would like to thank the Organizing Committee of the International Conference on Mathematical Modeling in Physical Sciences 2012
 for their invitation to present our work.\\

                                                                           \vspace{5mm}
                                                                           
                                                                                                                         
                                                        \centerline{\large\sc References}
 
                                                                          \vspace{5mm}

\noindent [1] J. Havrda, F. Charvat, \ \emph{Kybernetika} {\bf 3}, \ 30 (1967)\\ 
\noindent [2] Z. Dar\'{o}czy, \ \emph{Inf. Comp./ Inf. Contr.} {\bf 16}, 36 (1970)\\
\noindent [3] C. Tsallis, \  \emph{J. Stat. Phys.} {\bf 52}, \ 479 \ (1988)\\
\noindent [4] C. Tsallis, \  \emph{Introduction to Nonextensive Statistical Mechanics: Approaching a Complex\\
                                \hspace*{4mm} World},  \ Springer (2009)\\
\noindent [5] M. Costeniuc, R.S. Ellis, H. Touchette, B. Turkington, \ \emph{J. Stat. Phys.} {\bf 119}, 1283 (2005)\\ 
\noindent [6] H. Touchette, \  \emph{Europhys. Lett.} {\bf 96}, \ 50010 \ (2011)\\
\noindent [7] A. Campa, T. Dauxois, S. Ruffo, \ \emph{Phys. Rep.} {\bf 480}, \ 57 \ (2009)\\
\noindent [8] H. Touchette, \ Physica A {\bf 305}, \ 84 \ (2002)\\
\noindent [9] J. Zinn-Justin, \ \emph{Quantum Field Theory and Critical Phenomena}, \ 3rd Ed., \  Clarendon Press \\
                               \hspace*{4mm}  (1997)\\ 
\noindent [10] J. Glimm, A. Jaffe, \ \emph{Quantum Physics: A Functional Integral Point of View},  \ 2nd Ed., \\
                               \hspace*{6mm} Springer-Verlag \ (1987)\\ 
\noindent [11] J.W. Gibbs, \ \emph{Elementary Principles in Statistical Mechanics}, \ Yale University Press \ (1948)\\
\noindent [12] M. Veltmann, \ \emph{Quantum Theory of Gravitation}, \ in \ \emph{Methods in Field Theory}, \ Les Houches \\
                              \hspace*{6mm} 1975, \ Session XXVIII, \ R. Balian, J. Zinn-Justin, Eds., \ North-Holland \ (1976)\\
\noindent [13] A. Ashtekar, \ \emph{Introduction to Loop Quantum Gravity}, \ {\sf arXiv:1201.4598}\\
\noindent [14] R. Sorkin, \ \emph{J. Phys. Conf. Ser.} {\bf 174}, \ 012018 \ (2009)\\
\noindent [15] J. Ambjorn, A. Goerlich, J. Jurkiewicz, R. Loll, \ \emph{Nonperturbative Quantum Gravity}, \\
                              \hspace*{6mm}  {\sf arXiv:1203.3591}\\
\noindent [16] J. Polchinski, \ \emph{String Theory}, \  Vols,1, 2. \ Cambridge University Press (2005)\\
\noindent [17] J. Maldacena, \ \emph{The gauge/gravity duality}, \ {\sf arXiv:1106.6073}\\
\noindent [18] R. Sorkin, \ \emph{ The Statistical Mechanics of Black Hole Thermodynamics}, \ in \ \emph{Black Holes 
                             \hspace*{6mm} and Relativistic Stars}, \ R.M. Wald, \ Ed., \ University of Chicago Press (1998)\\
\noindent [19] R.M. Wald, \ \emph{Living Rev. Rel.} {\bf 4}, \  6 \ (2001)\\
\noindent [20] N. Kalogeropoulos, \ \emph{Physica A} {\bf 391}, \ 1120 \ (2012)\\
\noindent [21] N. Kalogeropoulos, \ \emph{Physica A} {\bf 391}, \ 3435 \ (2012)\\
\noindent [22] N. Kalogeropoulos, \emph{Vanishing largest Lyapunov exponent and Tsallis entropy}, {\sf arXiv:1203.2707}\\ 
\noindent [23] C. Beck, F. Schl\"{o}gl, \ \ \emph{Thermodynamics of chaotic systems: an introduction}, \  \   Cambridge\\
                             \hspace*{6mm} University Press \ (1993)\\
\noindent [24] N. Kalogeropoulos, \  \emph{Escort distributions and Tsallis entropy}, \   {\sf arXiv:1206.5127}\\
\noindent [25] A. J. Creaco, N. Kalogeropoulos, \ \emph{Nilpotence in Physics: the case of Tsallis entropy}\\
                            \hspace*{6mm} {\sf arXiv:1209.4180}\\ 
\noindent [26] L. Nivanen, A. Le Mehaut\'{e}, Q.A. Wang, \ \emph{Rep. Math. Phys.} {\bf 52}, \ 437 \ (2003)\\ 
\noindent [27] E.P. Borges, \ \emph{Physica A} {\bf 340}, \ 95 \ (2004)\\
\noindent [28] S. Abe, \  \emph{Phys. Lett. A} {\bf 271}, \ 74 \ (2000)\\
\noindent [29] R.J.V. Santos, \ \emph{J. Math. Phys.} {\bf 38}, \ 4104 \ (1997)\\
\noindent [30] H. Suyari, \ \emph{IEEE Trans. Inf. Theor.} {\bf 50}, \ 1783 \ (2004)\\
\noindent [31] A. Knapp, \ \emph{Lie Groups: Beyond an Introduction}, \ 2nd Ed., \ Birkh\"{a}user (2002)\\
\noindent [32] A. Shapere, F. Wilczek, \ \emph{Geometric Phases in Physics}, \ World Scientific \ (1989)\\
\noindent [33] R. Montgomery, \ \emph{A Tour of Subriemannian Geometries, Their Geodesics and Applications,} \\
                             \hspace*{6mm} Amer. Math. Soc.  \ (2006)\\
\noindent [34] M. Gromov, \ \emph{Carnot-Carath\'{e}odory spaces seen from within}, \ in \ \emph{Sub-Riemannian Geometry} \\ 
                             \hspace*{6mm}  A. Bella\"{i}che, J.-J. Risler, \ Eds., \ Birkh\"{a}user \ (1996)\\
\noindent [35] B. Mandelbrot, \ \emph{Ann. New York Acd. Sci.} {\bf 357}, \ 249 \ (1980)\\ 
\noindent [36] K. Falconer, \ \emph{Fractal Geometry: Mathematical Foundations and Applications}, \ John Wiley\\
                             \hspace*{6mm}  \& Sons \ (2003)\\
\noindent [37] W. Hurewicz, H. Wallman, \ \emph{Dimension Theory}, \ Princeton University Press (1948)\\
\noindent [38] M. Gromov, \ \emph{Metric Structures for Riemannian and non-Riemannian Spaces}, \ Birkh\"{a}user \\
                             \hspace*{6mm} (1999)\\
\noindent [39] M. Gromov, \ \emph{Publ. Math. I.H.E.S.} {\bf 53}, \ 53 \ (1981)\\ 
\noindent [40] M. Gromov, \ \emph{Asymptotic invariants of infinite groups}, in  \emph{Geometric Group Theory, Vol. 2} \\
                               \hspace*{6mm} G.A. Niblo, M.A. Roller (Eds.), \  Cambridge University Press (1993)\\ 
\noindent [41] C. Drutu, \ \emph{Int. J. Algebra Comput.} {\bf 12}, \ 99 \ (2002)\\ 
\noindent [42] H. Federer, \  \emph{Geometric Measure Theory}, \ Springer (1969)\\
\noindent [43] J. Cheeger, \ \emph{Geom. Funct. Anal.} {\bf 9}, \ 428 \ (1999)\\
\noindent [44] B. Kleiner, J. Mackay, \ \emph{Differentiable Structures on Metric Measure Spaces: a Primer}, \\
                              \hspace*{6mm}  {\sf arXiv:1108.1324}\\
\noindent [45] P. Pansu, \  \emph{Ann. Math.} {\bf 129}, \ 1 \ (1989)\\   
\noindent [46] S. Abe, \ \emph{Phys. Lett. A} {\bf 224}, \ 326 \ (1997)\\
\noindent [47] J. Milnor, \ \emph{J. Diff. Geom.} {\bf 2}, \ 447 \ (1968)\\ 
\noindent [48] J.A. Wolf, \ \emph{J. Diff. Geom.} {\bf 2}, \ 421 \ (1968)\\  
\noindent [49] J. Tits, \ \emph{Classification of algebraic semisimple groups}, \ in \   \emph{Algebraic Groups \\
                               \hspace*{6mm} and Discontinuous Subgroups}, \ Proc. Symp. Pure Math. {\bf 9}, \ AMS \ (1966)\\  
\noindent [50] B. Kleiner, \ \emph{J. Amer. Math. Soc.} {\bf 23}, \ 815 \ (2010)\\ 
\noindent [51] Y. Shalom, T. Tao, \ \emph{Geom. Funct. Anal.} {\bf 20}, 1502 \ (2010)\\
\noindent [52] C. Tsallis, M. Gell-Mann, Y. Sato, \ \emph{Proc. Natl. Acad. Sci.} {\bf 102}, \ 15377 \ (2005)\\
\noindent [53] R. Hanel, S. Thurner, \ \emph{Europhys. Lett.} {\bf 96}, \ 50003 \ (2011)\\ 
\noindent [54] S. Semmes, \ \emph{Rev. Mat. Iberoamericana} {\bf 12}, \ 337 \ (1996)\\
\noindent [55] M. Gromov, \ \emph{Hyperbolic groups}, \ in \ \emph{Essays in group theory}, \ S. Gersten (Ed.), \\ 
                           \hspace*{6mm}  MSRI Publ. {\bf 8}, \ Springer \  (1987)\\ 

\end{document}